\documentclass[a4paper,11pt]{article}
\pdfoutput=1

\usepackage{jcappub}
\renewcommand{\afterAuthorSpace}{\vskip 8pt}
\renewcommand{\afterAffiliationSpace}{\vskip 0pt}

\usepackage[T1]{fontenc} 
\usepackage{graphicx}
\usepackage{float}
\usepackage{placeins}
\usepackage{tablefootnote}
\usepackage{pbox}
\usepackage{booktabs}
\usepackage{makecell}
\usepackage{multirow}
\usepackage{array}
\usepackage{mleftright}
\usepackage[ISO]{diffcoeff}
\usepackage{siunitx}
\usepackage{hyperref}
\usepackage[multiple]{footmisc}
\usepackage[capitalise]{cleveref}

\makeatletter
\input{aas_macros.sty}
\let\jnl@style=\relax
\makeatother

\DeclareSIUnit{\pc}{pc}
\DeclareSIUnit{\kpc}{\kilo\pc}
\DeclareSIUnit{\Mpc}{\mega\pc}
\DeclareSIUnit{\Gpc}{\giga\pc}
\DeclareSIUnit{\hHubble}{\text{\ensuremath{h}}}
\crefname{section}{\S}{\S}
\crefformat{section}{\S#2#1#3}
\crefmultiformat{section}{\S#2#1#3}{ and~\S#2#1#3}{, \S#2#1#3}{ and~\S#2#1#3}
\AddToHook{cmd/appendix/before}{%
    \crefalias{chapter}{appendix}%
    \crefalias{section}{appendix}%
    \crefalias{subsection}{appendix}%
}

\title{Odd Loops: What Structures Can Produce a Parity-Violating Inflationary Trispectrum?}
\author[1]{William Ortolá Leonard}
\author[2]{and Zachary Slepian}
\affiliation[1]{Department of Physics, University of Florida,\\2001 Museum Rd., Gainesville, FL 32611, USA}
\affiliation[2]{Department of Astronomy, University of Florida,\\211 Bryant Space Science Center, Gainesville, FL 32611, USA}

\emailAdd{wortola@ufl.edu}
\emailAdd{zslepian@ufl.edu}

\abstract{Non-Gaussian perturbations generated during inflation can often be described perturbatively via integrals of kernels acting on Gaussian fields. In this work, we investigate, in a model-independent manner, whether these kernels can produce parity-violating (PV) contributions to the trispectrum, and which Wick's theorem contraction configurations of the Gaussian fields lead to them. In detail, we evaluate all possible contraction configurations arising from the third- and second-order density perturbations. We find that, for the third-order perturbations, only one configuration can support PV contributions, as it retains three independent wave-vectors. In contrast, none of the four second-order perturbation contraction configurations can produce a PV contribution under statistical isotropy. These results offer a first step toward deriving kernels from inflationary models capable of generating a PV trispectrum. Such kernels could then be used to perform PV searches with current and upcoming galaxy redshift surveys such as DESI, Roman, SPHEREx, and Euclid, and photometric efforts such as LSST. Our results also offer a guide for exploration of PV with CMB data-sets such as Planck.}

\newcommand{\vecphi}{\vec{\varphi}}
\newcommand{\veck}{\mathbf{k}}
\newcommand{\vecq}{\mathbf{q}}
\newcommand{\vecp}{\mathbf{p}}

\newcommand{\hatk}{\widehat{\mathbf{k}}}

\newcommand{\PP}{\mathcal{P}}

\newcommand{\dc}[1]{\delta_L({#1})}
\newcommand{\dirac}{\delta^{[3]}_{\rm D}}

\newcommand{\Fthree}[3]{F^{(3)}({#1},{#2},{#3})}
\newcommand{\Ftwo}[2]{F^{(2)}({#1},{#2})}
\newcommand{\Rthree}[1]{\tau_{1113}^{{#1}}(\veck_1,\veck_2,\veck_3,\veck_4)}
\newcommand{\Rtwo}[1]{\tau_{1122}^{{#1}}(\veck_1,\veck_2,\veck_3,\veck_4)}

\begin{document}

\maketitle

\section{Introduction}
Parity symmetry requires that physical observables remain unchanged under spatial inversion, $(x, y, z) \to (-x, -y, -z)$. In the Universe's large-scale structure (LSS), correlation functions (CFs) involving fewer than three independent spatial vectors cannot distinguish a configuration from its mirror image \cite{cahn_parity}. The lowest-order CF described by three spatial vectors is the 4-Point Correlation Function (4PCF). The 4PCF, which measures excess clustering of quadruplets over and above random, quantifies tetrahedra, and thus its handedness can be characterized by the scalar triple product of the three separation vectors (see Figure 1 of \cite{hou_parity}). Consequently, the 4-Point Correlation Function (4PCF), and its Fourier-space counterpart, the trispectrum, are the lowest-order \textit{scalar} clustering statistics capable of probing parity violation (PV) in LSS \cite{cahn_parity}.\footnote{LSS is intrinsically a 3D field. As shown in \cite{Ortola_2d_Test_PV}, using 2D fields, the 3PCF is the corresponding lowest-order CF capable of detecting PV in the Universe.}\textsuperscript{,}\footnote{\cite{jeong_fossil} proposed for 3D LSS a related statistic to the 4PCF; \cite{shiraishii} proposed to use the trispectrum for the CMB. \cite{CMB_Greco_PV,CMB_PV_Test_Philcox,CMB_Philcox_Proposal} have also explored PV in the CMB.}

Observationally, measurements of the PV galaxy 4PCF from Sloan Digital Sky Survey Baryon Oscillation Spectroscopic Survey (SDSS BOSS) reported evidence for PV in the clustering of LSS \cite{hou_parity,phil_parity}. However, in a subsequent study of the same dataset but with a different statistical technique, \cite{Krolewski_No_PV_det} found between 0 and 2.5$\sigma$ evidence for PV, depending on the analysis configuration used. Similar results have been found in the Luminous Red Galaxy (LRG) sample of the Dark Energy Spectroscopic Instrument (DESI) Data Release 1. \cite{Slepian_Parity_Meas_DESI,Hou_Parity_odd_2} found evidence of PV in their auto-correlation analysis---which captures the PV 4PCF within individual survey regions. \cite{Slepian_Parity_Meas_DESI,Hou_Parity_odd_2} found no evidence for such a signal in their cross-correlation analyses, which examine correlations between PV 4PCF measurements from different survey regions. These observational developments motivate a more detailed \textit{theoretical} understanding of how PV contributions may appear in the 4PCF. Alternative approaches to searching for PV have also been developed. \cite{jamieson_pops} introduced the parity-odd power (POP) spectra, which compress the parity-odd trispectrum into power-spectrum-like observables. Similarly, \cite{GAO_QUIJOTE_Kurto_Test} developed parity-odd kurto spectra and tested them using the \textsc{Quijote} and \textsc{Quijote-Odd} simulations \cite{villaescusa2020quijote,coulton}. Subsequently, \cite{GAO_Kurto_TEST_DESI_BOSS} applied the kurto spectra to the BOSS and DESI Y1 LRG samples, finding no evidence for PV in either survey. More recently, \cite{Ortola_2d_Test_PV} proposed a new method to search for PV using the galaxy 3PCF restricted to 2D spherical shells, while \cite{Yap_Bispectrum_PV} introduced a complementary approach using the density-density-momentum bispectrum to probe primordial PV. 

On the theoretical side, different mechanisms have been proposed through which new physics in the early Universe could generate a PV primordial scalar trispectrum \cite{reinhard_axion,Caprini_helicity_Inflation_PV_model,shiraishii,Bartolo_PV_CMB,Reinhard_Polyspectra_Axion,cabass_no_go,cabass,stefanyszyn,thavanesan_no_go,Sadra_PV_inflation}. Using \textsc{Quijote-Odd} simulations, \cite{coulton} demonstrated that primordial PV signatures can survive nonlinear gravitational evolution and remain detectable in the late-time \textit{matter} trispectrum. However, \cite{Ortola_Quijote_No_PV_Halos} found that no PV signal is detected at the \textit{halo} level in the same \textsc{Quijote-Odd} simulations, and suggests that this may be due to shot noise. 

In most models of inflation that produce PV, one proceeds by expanding in small perturbations around the uniform background field, and then obtaining a perturbative solution of the differential equations governing the model; for instance \cite{reinhard_axion,Caprini_helicity_Inflation_PV_model,shiraishii,Bartolo_PV_CMB,Reinhard_Polyspectra_Axion,Abolhasani_Anisotropies_EarlyUniverse_EFT}. These solutions generically have the structure of a kernel integrated against two or more copies of the linear density field, which is taken to be a Gaussian Random Field (GRF). An important question thus arises: which kernels can lead to a PV trispectrum? In this work, we endeavor to answer this question while remaining as agnostic as possible to the detailed forms of the kernels. This allows us to make general statements about what type of structure can lead to a PV loop contribution to the primordial trispectrum.

To perform our exploration, we follow the same strategy as in Eulerian Standard Perturbation Theory (SPT). In SPT, the density and velocity fields are expanded perturbatively through mode-coupling kernels $F^{(n)}$ and $G^{(n)}$ \cite{Bernardeau_2002}. Within this framework, the leading-order non-Gaussian trispectrum\footnote{\cite{Ortola_4PCF,gualdi_2021_trispec} name this term the tree-level trispectrum. However, in Effective Field Theory and Quantum Field Theory these terms of the trispectrum might be named the one-loop contributions. To avoid confusion in this work, we use the terminology ``leading-order non-Gaussian contributions".}receives contributions from third-order density perturbations, $\langle\delta_L\delta_L\delta_L\delta^{(3)}\rangle$, and from pairs of second-order perturbations, $\langle\delta_L\delta_L\delta^{(2)}\delta^{(2)}\rangle$ \cite{Ortola_4PCF}, where subscript $L$ denotes a linear density field. Analogously, for the primordial PV perturbative solutions we can form the primordial trispectrum as
\begin{align}
&(2\pi)^3 \,\dirac(\veck_{1234})\, T(\veck_1,\veck_2,\veck_3,\veck_4) = \,\Bigg\langle \delta(\mathbf{k}_1)\;\delta(\mathbf{k}_2)\;\delta(\mathbf{k})\;\delta(\mathbf{k})\Bigg\rangle \\
& = \Bigg\langle \dc{\veck_1}\dc{\veck_2}\Bigg\rangle\Bigg\langle \dc{\veck_3}\dc{\veck_4}\Bigg\rangle + {\rm perms.} \nonumber \\
&\;+  \Bigg[\Bigg\langle \dc{\veck_1}\,\dc{\veck_2}\,\dc{\mathbf{k}_3} \,\delta^{(3)}(\veck_4)\Bigg\rangle + \Bigg\langle \dc{\veck_1}\,\dc{\veck_2}\,\delta^{(2)}{\mathbf{k}_3} \,\delta^{(2)}(\veck_4)\Bigg\rangle + {\rm perms.}\Bigg].\nonumber
\end{align}
The first term in the second equality is the disconnected piece of the trispectrum that is composed of four linear density fields, which can be decomposed in products of two power spectra via Wick's theorem \cite{Wick_Theorem}. The remaining terms are the PV leading-order trispectrum. In this work we will be evaluating 
\begin{align}
&\Rthree{} \equiv \Bigg\langle \dc{\veck_1}\,\dc{\veck_2}\,\dc{\mathbf{k}_3} \,\delta^{(3)}(\veck_4)\Bigg\rangle \nonumber \\
\\
&\Rtwo{} \equiv \Bigg\langle \dc{\veck_1}\,\dc{\veck_2}\,\delta^{(2)}({\mathbf{k}_3}) \,\delta^{(2)}(\veck_4)\Bigg\rangle .\nonumber
\end{align}

In \Cref{sec:3rd_kernel_Eval} of this work, we investigate all possible configurations that would survive and give rise to a PV trispectrum from the third-order densities; in \Cref{sec:2nd_kernel_Eval}, we examine all the possible configurations from the second-order densities. We offer concluding discussion in  \Cref{Sec:Discussion}. 

\section{Third-Order Kernel Evaluation}\label{sec:3rd_kernel_Eval}
\begin{figure}
    \centering
    \includegraphics[width=1.0\linewidth]{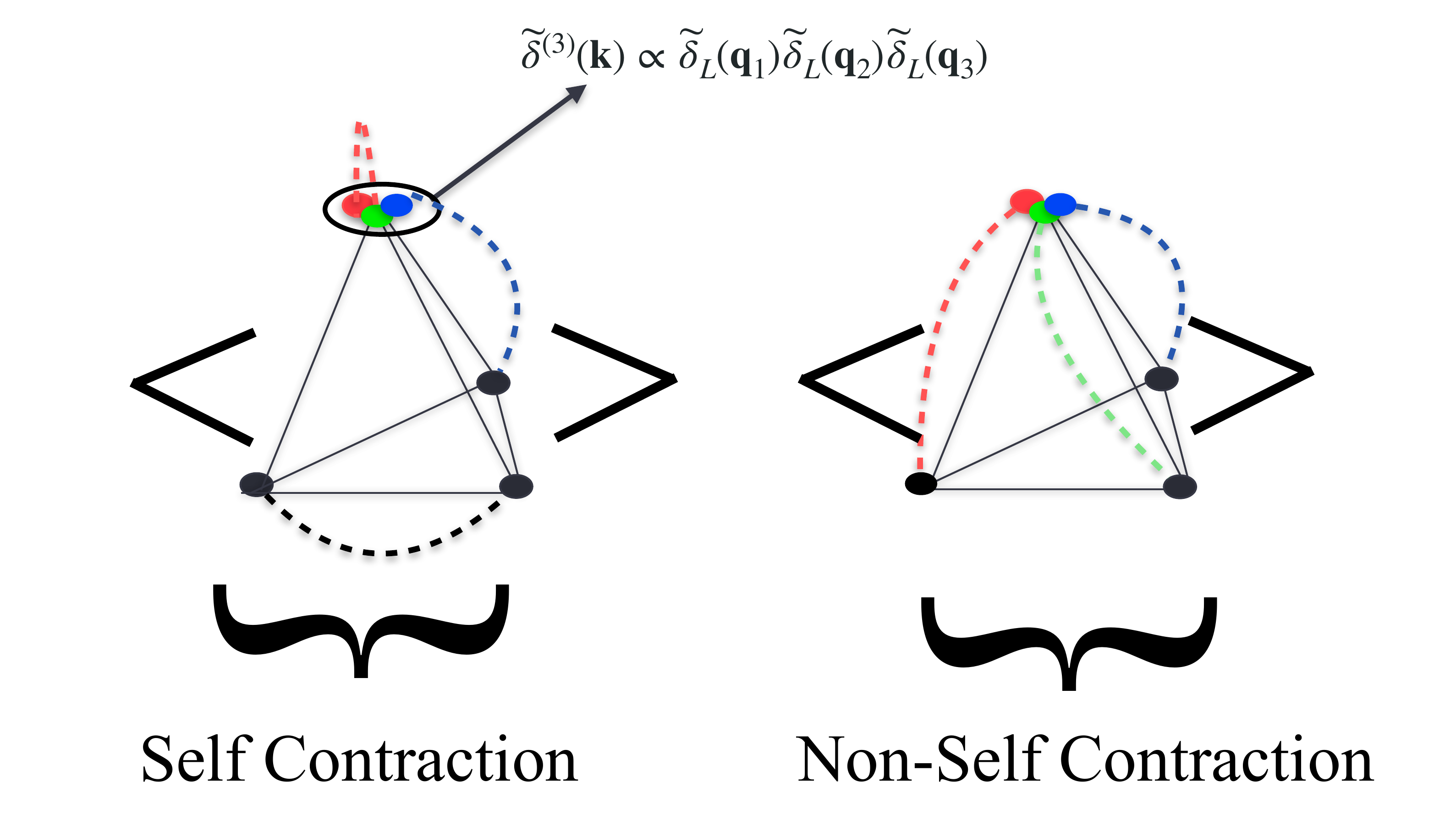}
    \caption{Schematic representation of the two possible Wick-contraction configurations arising from the third-order contribution to the tree-level trispectrum, $\langle \delta_L\delta_L\delta_L\delta^{(3)}\rangle$. The colored dots represent the three linear density fields composing the third-order density perturbation, while the black dots represent the linear density fields. The dashed lines indicate the Wick contractions. Left: Self-Contraction (SC) configuration, in which two linear fields within the third-order perturbation contract with one another. Right: Non-Self-Contraction (NSC), in which each of the three linear fields contracts with a different external linear density field. Only the NSC configuration contributes to the connected tree-level trispectrum. We follow the same diagrammatic conventions as in \cite{ortola_cov_I,ortola_cov_II}.}
    \label{fig:third_order_configurations}
\end{figure}

In this section, we evaluate the third-order kernel and determine the possible configurations that give rise to PV contributions. We begin by expanding the third-order density contrast in terms of the third-order kernel as
\begin{align}\label{eq:T3_def}
&\Rthree{} = \int d^3 \vecq_1\,d^3 \vecq_2\,d^3 \vecq_3 \;\Fthree{\vecq_1}{\vecq_2}{\vecq_3}\dirac
\left( \mathbf{q}_1+\mathbf{q}_2+\mathbf{q}_3-\mathbf{k}_4 \right)\nonumber\\
&\qquad\qquad\qquad\qquad\quad \times 
\Bigg\langle \dc{\veck_1}\,\dc{\veck_2}\,\dc{\mathbf{k}_3}\dc{\mathbf{q}_1}\;\dc{\mathbf{q}_2}\;\dc{\mathbf{q}_3}\Bigg\rangle .
\end{align}
Next, we apply Wick's theorem to contract the linear fields in the last line of \Cref{eq:T3_def} and evaluate the two possible configurations shown in \Cref{fig:third_order_configurations}.

\subsection{Evaluation of the Self-Contraction Configuration}\label{sec:SC_eval}
Using Wick's theorem on \Cref{eq:T3_def} for the Self-Contraction (SC) configuration shown in \Cref{fig:third_order_configurations} results in 
\begin{align}
&\Rthree{\rm (SC)} = \int d^3 \vecphi\int d^3 \vecq_1\,d^3 \vecq_2\,d^3 \vecq_3 \;\Fthree{\vecq_1}{\vecq_2}{\vecq_3}\dirac
\left( \mathbf{q}_1+\mathbf{q}_2+\mathbf{q}_3-\mathbf{k}_4 \right)\nonumber\\
&\qquad\qquad\qquad\qquad\times 
\Bigg\langle \dc{\veck_1}\dc{\veck_2}\Bigg\rangle\Bigg\langle\dc{\mathbf{k}_3}\dc{\mathbf{q}_3}\Bigg\rangle\Bigg\langle\dc{\mathbf{q}_1}\dc{\mathbf{q}_2}\Bigg\rangle  + {\rm 5 \;perms.},
\end{align}
where $5$ perms. indicates the other five combinations\footnote{We find the number of permutations in three steps. i) We choose one of the third-order densities, then pair this third-order density with another one of the third-order densities; one has two possibilities. ii) We pair the remaining third-order density with one of the linear densities; one has three possibilities. iii) We pair the remaining linear densities; there is one possibility. The total number of permutations = 3 $\times$ 2$\times$ 1  = 6 perms. Throughout this work, we calculate the number of permutations using the same procedure as in \cite{ortola_cov_I,ortola_cov_II}.} from Wick's theorem that we obtain for this SC structure. For brevity, we do not write this out in what follows, but we intend that it should be implicitly in our calculations. We may then evaluate the ensemble averages as power spectra and obtain
\begin{align}
&\Rthree{\rm (SC)} = (2\pi)^{9} \dirac(\veck_1+\veck_2) P(k_1) P(k_3) \nonumber \\
& \qquad\times \int d^3 \vecphi\int d^3 \vecq_1 d^3 \vecq_2 d^3 \vecq_3\; \Fthree{\vecq_1}{\vecq_2}{\vecq_3} \dirac
\left( \mathbf{q}_1+\mathbf{q}_2+\mathbf{q}_3-\mathbf{k}_4 \right) \nonumber \\
& \qquad\times \dirac(\veck_3+\vecq_3) \dirac(\vecq_1+\vecq_2)\;P(q_1).
\end{align}
Then, we can integrate over $\vecq_2$ and $\vecq_3$ to simplify our result as
\begin{align}\label{eq:SC_conf}
&\Rthree{\rm (SC)} = P(k_1) P(k_4) \dirac(\veck_1+\veck_2)\dirac(\veck_3+\veck_4) \nonumber \\
& \qquad \qquad \qquad \qquad \quad \times \int d^3 \vecq_1\; \Fthree{\vecq_1}{-\vecq_1}{\veck_4} \;P(q_1).
\end{align}
\Cref{eq:SC_conf} shows that the final angular dependence will be only that of the wave-vector $\veck_4$. Due to isotropy, we cannot have any dependence on the absolute directions of the $\hatk_i$, which means that the $\veck_4$ dependence will disappear and this self-contracting contribution can only be parity even.

\subsection{Evaluation of the Non-Self-Contraction Configuration}\label{sec:NSC_eval}
Next, we apply Wick's theorem to perform contractions on \Cref{eq:T3_def} for the Non-Self-Contraction (NSC) configuration shown in \Cref{fig:third_order_configurations}, for which we obtain
\begin{align}
&\Rthree{\rm (NSC)} =  \int d^3 \vecq_1\,d^3 \vecq_2\,d^3 \vecq_3 \;\Fthree{\vecq_1}{\vecq_2}{\vecq_3}\dirac
\left( \mathbf{q}_1+\mathbf{q}_2+\mathbf{q}_3-\mathbf{k}_4 \right)\\
&\qquad\qquad\qquad\qquad\quad \times 
\Bigg\langle \dc{\veck_1}\dc{\vecq_2}\Bigg\rangle\Bigg\langle\dc{\mathbf{k}_2}\dc{\mathbf{q}_2}\Bigg\rangle\Bigg\langle\dc{\mathbf{k}_3}\dc{\mathbf{q}_3}\Bigg\rangle +{\rm 5\;perms.}\footnotemark[5]\nonumber
\end{align}
\footnotetext[5]{We Find the number of permutations in three steps. i) We choose one of the third-order densities, then pair this third-order density with one of the linear densities; one has three possibilities. ii) We pair another third-order density with one of the two-leftover linear densities; one has two possibilities. iii) We pair the remaining third-order density with the remaining linear densities; there is one possibility. The total number of permutations = 3 $\times$ 2$\times$ 1  = 6 perms.}Expressing the ensemble averages in terms of their power spectra and evaluating all the $\vecq_i$ integrals results in 
\begin{align}
&\Rthree{\rm (NSC)} = (2\pi)^{9} P(k_1) P(k_2) P(k_3)\,\dirac(\veck_1+\veck_2+\veck_3+\veck_4) \nonumber\\
& \qquad \qquad \qquad \qquad \quad\, \times \Fthree{-\veck_1}{-\veck_2}{-\veck_3} .
\end{align}
\cite{Ortola_4PCF} demonstrated in the SPT framework that the perturbative kernels can be expanded, decomposed and reorganized such that the angular and radial dependence is separated. Applying the same idea to the above expression, we can rewrite the third-order kernel in terms of the isotropic basis functions 
\begin{align}
&\Rthree{\rm (NSC)} = (2\pi)^{9} P(k_1) P(k_2) P(k_3)\,\dirac(\veck_1+\veck_2+\veck_3+\veck_4) \nonumber\\
& \qquad \qquad \qquad \qquad \quad\, \times \sum_{\ell_1,\ell_2,\ell_3} (-1)^{\ell_1,\ell_2,\ell_3} \,\zeta_{\ell_1,\ell_2,\ell_3} (k_1,k_2,k_3) \PP_{\ell_1,\ell_2,\ell_3} (\hatk_1,\hatk_2,\hatk_3).
\end{align}
The above result shows that the NSC configuration is sensitive to PV multipoles and in principle any early-universe parity-violating physics could give rise to such PV contributions of the kernel. 

\section{Second-Order Kernel Evaluation} \label{sec:2nd_kernel_Eval}
\begin{figure}
    \centering
    \includegraphics[width=1.0\linewidth]{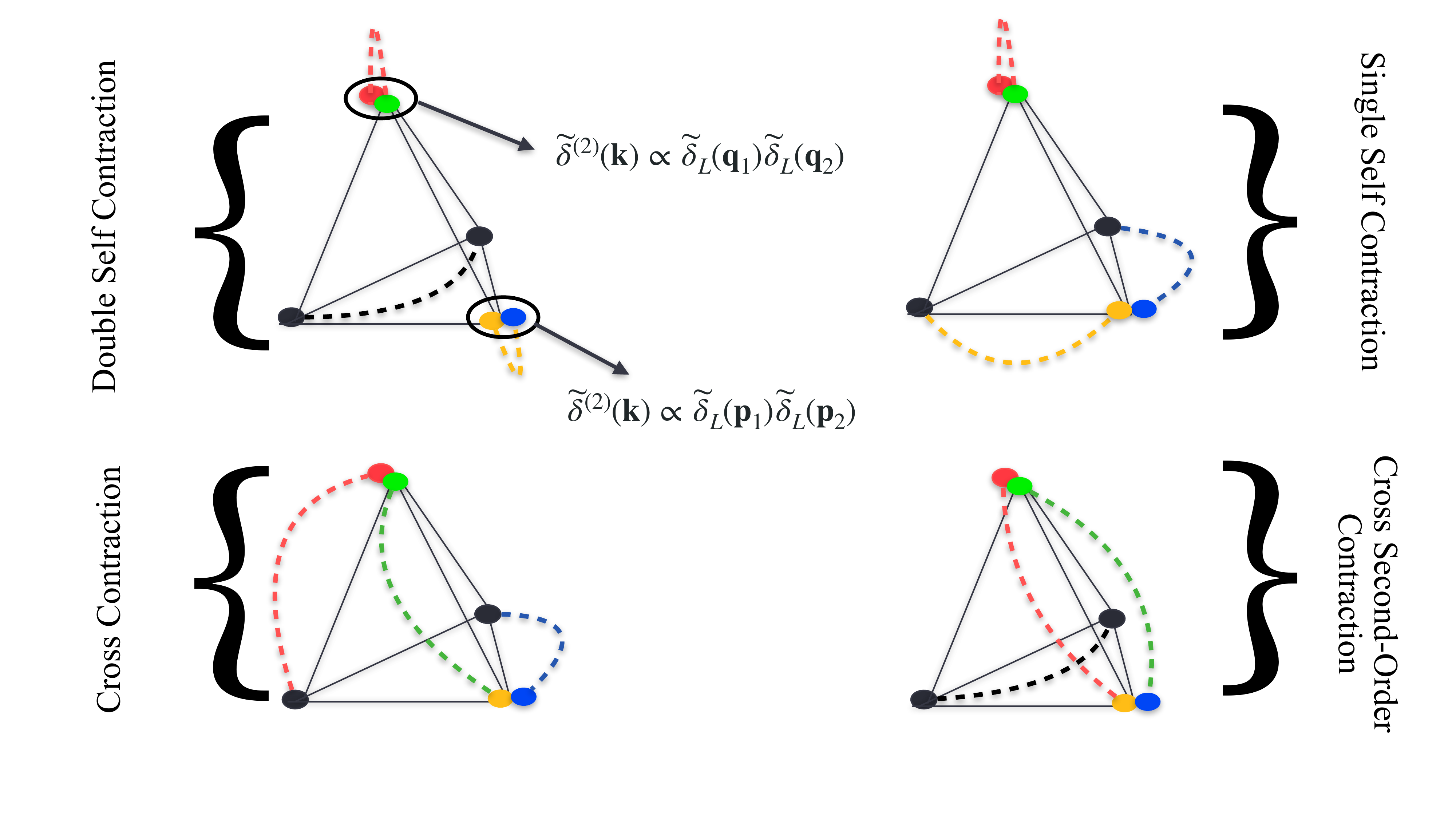}
    \caption{Schematic representation of the four possible Wick-contraction configurations arising from the second-order contribution to the tree-level trispectrum, $\langle \delta_L\delta_L\delta^{(2)}\delta^{(2)}\rangle$. The colored dots represent the linear density fields composing each second-order density perturbation, while the dashed lines indicate the Wick contractions. The configurations are classified as Double Self-Contraction (DSC), where both second-order perturbations contract internally; Single Self-Contraction (SSC), where only one contracts internally; Cross Contraction (CC), where the two second-order perturbations are connected to each other and to the external linear fields; and Cross Second-Order Contraction (CSOC), where the two second-order perturbations contract exclusively with one another. None of these configurations can contribute a PV SPT kernel. We follow the same diagrammatic conventions as in \cite{ortola_cov_I,ortola_cov_II}.}
    \label{fig:second_order_configurations}
\end{figure}

In this section, we evaluate the second-order kernel and determine the possible configurations that give rise to a PV contribution. We begin by expanding the second-order density contrast in terms of the second-order kernel as
\begin{align}\label{eq:T2_def}
&\Rtwo{} = \int d^3 \vecphi\int d^3 \vecq_1\,d^3 \vecq_2 \;\Ftwo{\vecq_1}{\vecq_2}\dirac
\left( \mathbf{q}_1+\mathbf{q}_2-\mathbf{k}_3 \right)\nonumber\\
&\qquad\qquad\qquad \qquad\,\,\,\,\times  \int d^3 \vecp_1\,d^3 \vecp_2 \;\Ftwo{\vecp_1}{\vecp_2}\dirac
\left( \mathbf{p}_1+\mathbf{p}_2-\mathbf{k}_4 \right)\nonumber\\
&\qquad\qquad\qquad\qquad\;\; \times 
\Bigg\langle \dc{\veck_1}\,\dc{\veck_2}\,\dc{\mathbf{q}_1}\;\dc{\mathbf{q}_2}\;\dc{\vecp_1}\dc{\vecp_2}\Bigg\rangle.
\end{align}
Next, we will be applying Wick's contractions to the last line of \Cref{eq:T2_def} to evaluate the three possible configurations shown in \Cref{fig:second_order_configurations}.

\subsection{Evaluation of the Double Self-Contraction Configuration}
Using Wick's theorem on \Cref{eq:T2_def} for the Double Self-Contraction (DSC) configuration shown in \Cref{fig:second_order_configurations} results in 
\begin{align}
& \Rtwo{\rm (DSC)}=\int d^3\vecphi\int d^3 \vecq_1\,d^3 \vecq_2 \;\Ftwo{\vecq_1}{\vecq_2}\dirac
\left( \mathbf{q}_1+\mathbf{q}_2-\mathbf{k}_3 \right)\nonumber\\
&\qquad\qquad\qquad \qquad\,\,\,\,\times  \int d^3 \vecp_1\,d^3 \vecp_2 \;\Ftwo{\vecp_1}{\vecp_2}\dirac
\left( \mathbf{p}_1+\mathbf{p}_2-\mathbf{k}_4 \right)\\
&\qquad\qquad\qquad\qquad\;\; \times 
\Bigg\langle \dc{\veck_1}\dc{\veck_2}\Bigg\rangle\Bigg\langle\dc{\mathbf{q}_1}\dc{\mathbf{q}_2}\Bigg\rangle\Bigg\langle\dc{\vecp_1}\dc{\vecp_2}\Bigg\rangle \nonumber.
\end{align}
This configuration only has one permutation as we require that both sets of second-order densities be contracted with themselves. Evaluating the ensemble average products as power spectra and evaluating the $\vecq_2$ and $\vecp_2$ integrals we obtain
\begin{align}
& \Rtwo{\rm (DSC)}=(2\pi)^{9}\,\dirac
\left( \mathbf{k}_3 \right)\,\dirac
\left( \mathbf{k}_4 \right)\,\dirac(\veck_1+\veck_2)\nonumber\\
& \qquad \qquad \qquad\qquad \quad \; \times P(k_1)\left[\int d^3 \vecq_1 \;\Ftwo{\vecq_1}{-\vecq_1}\,P(q_1)\right]^2.\nonumber\\
\end{align}
The result above squares the second-order kernel prohibiting any parity violating signal from manifesting. We note that above, we have exploited the fact that $\vecq_1$ and $\vecp_1$ are dummy variables that are integrated over to write the product in between square parenthesis raised to the second power. 

\subsection{Evaluation of the Single Self-Contraction Configuration}
Next, we use Wick's theorem on \Cref{eq:T2_def} for the Single Self-Contraction (SSC) configuration shown in \Cref{fig:second_order_configurations}. We find
\begin{align}
& \Rtwo{\rm (SSC)}=\int d^3 \vecq_1\,d^3 \vecq_2 \;\Ftwo{\vecq_1}{\vecq_2}\dirac
\left( \mathbf{q}_1+\mathbf{q}_2-\mathbf{k}_3 \right)\nonumber\\
&\qquad\qquad\qquad \qquad\,\,\,\,\times  \int d^3 \vecp_1\,d^3 \vecp_2 \;\Ftwo{\vecp_1}{\vecp_2}\dirac
\left( \mathbf{p}_1+\mathbf{p}_2-\mathbf{k}_4 \right)\nonumber\\
&\qquad\qquad\qquad\qquad\;\; \times 
\Bigg\langle \dc{\veck_1}\dc{\vecq_1}\Bigg\rangle\Bigg\langle\dc{\mathbf{k}_2}\dc{\mathbf{q}_2}\Bigg\rangle\Bigg\langle\dc{\vecp_1}\dc{\vecp_2}\Bigg\rangle + \;3\;{\rm perm.}\footnotemark[6]
\end{align}
\footnotetext[6]{We find the number of permutations in three steps. i) We choose one of the second-order densities, then pair this second-order density with one of the linear densities; one has two possibilities. ii) We pair the other second-order density with the two-leftover linear densities; there is one possibility. iii) We repeat i) and ii) but chosing the other set of second order densities; one has two possibilities. The total number of permutations = 2 $\times$ 1$\times$ 2  = 4 perms.}Evaluating the ensemble average products as power spectra and computing the $\vecq_1$, $\vecq_2$, and $\vecp_2$ integrals we obtain
\begin{align}
& \Rtwo{\rm (SSC)}=(2\pi)^9 \;\Ftwo{-\veck_1}{-\veck_2}\dirac
\left( \mathbf{k}_1+\mathbf{k}_2+\mathbf{k}_3 \right)\nonumber\\
&\qquad\qquad\qquad \qquad\,\,\,\,\times \dirac
\left(\mathbf{k}_4 \right) \int d^3 \vecp_1\;\Ftwo{\vecp_1}{-\vecp_1}\,P(p_1)
\end{align}
The above result is not compatible with a PV component since the second-order kernel inside the integral becomes a constant after integration, and the kernel outside of the integral can only produce dot products once we apply isotropy.

\subsection{Evaluation of the Cross-Contraction Configuration}
We also evaluate, using Wick's theorem on \Cref{eq:T2_def}, the Cross-Contraction (CC) configuration shown in \Cref{fig:second_order_configurations}. We obtain
\begin{align}
& \Rtwo{\rm (CC)}=\int d^3 \vecq_1\,d^3 \vecq_2 \;\Ftwo{\vecq_1}{\vecq_2}\dirac
\left( \mathbf{q}_1+\mathbf{q}_2-\mathbf{k}_3 \right)\nonumber\\
&\qquad\qquad\qquad \qquad\,\,\,\,\times  \int d^3 \vecp_1\,d^3 \vecp_2 \;\Ftwo{\vecp_1}{\vecp_2}\dirac
\left( \mathbf{p}_1+\mathbf{p}_2-\mathbf{k}_4 \right)\nonumber\\
&\qquad\qquad\qquad\qquad\;\; \times 
\Bigg\langle \dc{\veck_1}\dc{\vecq_1}\Bigg\rangle\Bigg\langle\dc{\mathbf{q}_2}\dc{\mathbf{p}_2}\Bigg\rangle\Bigg\langle\dc{\vecp_1}\dc{\veck_2}\Bigg\rangle+ \;3\;{\rm perm.}\footnotemark[7]
\end{align}
\footnotetext[7]{We find the number of permutations in three steps. i) We choose one of the second-order densities, then pair this second-order density with one of the linear densities; one has two possibilities. ii) We pair the other second-order density and pair it with one of the other two of second-order densities (in the other set of second-order densities); one has two possibilities. iii) We pair the left-over second-order density with the linear density; there is one possibility. The total number of permutations = 2 $\times$ 2$\times$ 1  = 4 perms.}Evaluating the ensemble average products as power spectra and computing all the intermediate integrals we obtain
\begin{align}
& \Rtwo{\rm (CC)}= (2\pi)^9 P(k_1)P(k_2) P(|\veck_1+\veck_3|)\;\Ftwo{-\veck_1}{\veck_1+\veck_3} \nonumber\\
&\qquad\qquad\qquad \qquad\,\,\,\,\times  \Ftwo{-\veck_2}{-\veck_1-\veck_3}\dirac
\left( \mathbf{k}_1+\mathbf{k}_2+\mathbf{k}_3 +\veck_4\right)
\end{align}
Once again isotropy only allows the formation of dot products, so even though we have different wave-vector combinations, we cannot form any parity odd contributions to the kernel. We note, however, that if the second-order kernel was anisotropic the above result would enable a PV contribution.  

\subsection{Evaluation of the Cross Second-Order Contraction Configuration}
Finally, we evaluate, using Wick's theorem on \Cref{eq:T2_def}, the Cross Second-Order Contraction (CSOC) configuration shown in \Cref{fig:second_order_configurations}, for which we obtain
\begin{align}
& \Rtwo{\rm (CSOC)}=\int d^3 \vecq_1\,d^3 \vecq_2 \;\Ftwo{\vecq_1}{\vecq_2}\dirac
\left( \mathbf{q}_1+\mathbf{q}_2-\mathbf{k}_3 \right)\nonumber\\
&\qquad\qquad\qquad \qquad\,\,\,\,\times  \int d^3 \vecp_1\,d^3 \vecp_2 \;\Ftwo{\vecp_1}{\vecp_2}\dirac
\left( \mathbf{p}_1+\mathbf{p}_2-\mathbf{k}_4 \right)\nonumber\\
&\qquad\qquad\qquad\qquad\;\; \times 
\Bigg\langle \dc{\veck_1}\dc{\veck_2}\Bigg\rangle\Bigg\langle\dc{\mathbf{q}_1}\dc{\mathbf{p}_1}\Bigg\rangle\Bigg\langle\dc{\vecp_2}\dc{\vecq_2}\Bigg\rangle + \; 1{\rm perm.}\footnotemark[8]
\end{align}
\footnotetext[8]{We find the number of permutations in three steps. i) We choose one of the second-order densities, then pair this second-order density with one of the other set of second-order densities; one has two possibilities. ii) We pair the other second-order density with one of the left-over second-order density; there is one possibility. iii) We pair the left-over linear densities; there is one possibility. The total number of permutations = 2 $\times$ 1$\times$ 1  = 2 perms.}Evaluating the ensemble average products as power spectra and computing all the intermediate integrals except $\vecq_2$ we obtain
\begin{align}\label{eq:CSOC_res}
& \Rtwo{\rm (CSOC)}= (2\pi)^9 P(k_1)\,\dirac(\veck_1 + \veck_2)\,\dirac
\left(\mathbf{k}_3 +\veck_4\right) \nonumber\\
&\qquad\qquad\qquad \qquad\,\,\,\,\times  \int d^3\vecq_2\, P(q_2) P(|\veck_3-\vecq_2|) \left[\Ftwo{\veck_3-\vecq_2}{\vecq_2}\right]^2
\end{align}
Several factors imply that \Cref{eq:CSOC_res} is parity-even only: i) The second-order kernel square cancel any PO contribution. ii) The only angular dependence will result from $\veck_3$ and by isotropy that dependence must vanish.

\section{Concluding Discussion} \label{Sec:Discussion}
In this work, we have investigated the configurations of the tree-level trispectrum that produce PV. Presuming that an inflationary model leading to PV has a perturbative solution given by integrals of kernels against two or more linear (\textit{i.e.} Gaussian) density fields, we have evaluated the different Wick-contraction configurations arising from the third- and second-order kernels. We obtain two configurations from the third-order contribution and four from the second order, as shown in \Cref{fig:third_order_configurations,fig:second_order_configurations}. 

For the third-order kernel, we find that the SC configuration, \Cref{sec:SC_eval}, cannot contribute to the PV trispectrum, since the resulting trispectrum depends on the absolute orientation of a single wave-vector, and such a dependence must vanish under isotropy. However, the NSC configuration, \Cref{sec:NSC_eval}, retains three independent wave-vectors and can, in principle, contain PV contributions. This is because from these three ave-vectors one can form the scalar triple product that a PV trispectrum requires. 

For the second-order kernels, we have evaluated all four possible configurations and found that none can produce a PO contribution under statistical isotropy. This is because the second-order kernels depend on only two wave-vectors and cannot form the scalar triple product required for parity violation \cite{cahn_parity,cahn_iso,iso_gen}. 

Therefore, our results suggest that, within the perturbative framework considered here, and assuming statistical isotropy, the third-order kernel is the lowest-order kernel capable of contributing to the PV leading-order non-Gaussian trispectrum.

An important future direction will be developing the perturbative approach to solve the equations of motion governing primordial perturbations in the presence of PV physics. For example, certain early-Universe models introduce a $U(1)$ gauge field that can source parity-violating contributions to the primordial perturbations \cite{reinhard_axion,Reinhard_Polyspectra_Axion,niu_axion_trispectrum}. By expanding the relevant fields perturbatively and solving their equations of motion recursively, it may be possible to derive a set of kernels analogous to those appearing in standard perturbation theory. Of particular interest would be determining the explicit form of the third-order kernel and whether it contains the dependence on three independent wave-vectors necessary to generate a PV trispectrum. Such a calculation would also allow us to investigate the time dependence of these kernels and whether their solutions can be separated into spatial and temporal components. Ultimately, these primordial contributions could be propagated through standard gravitational evolution to investigate their signatures in the late-time matter and galaxy 4PCF. Alternatively, the general form of the third-order kernel identified in this work as capable of generating a PV trispectrum could motivate a model-independent approach. By treating the coefficients of such a kernel as free parameters, we could construct theoretical templates that can be directly constrained using galaxy 4PCF measurements from galaxy surveys such as DESI, Roman, SPHEREx, Euclid, and LSST.

In conclusion, our results identify the perturbative configurations that can support PV contributions to the leading-order non-Gaussian trispectrum under the assumption of statistical isotropy. Although these results do not establish the explicit form or physical origin of a PV kernel, they provide a starting point for investigating how parity-violating physics in the early Universe could generate higher-order correlations that survive until late times. Deriving these kernels from the underlying equations of motion would provide a connection between early-Universe parity-violating mechanisms and the signatures that can be searched for in present-day cosmological observations.

\section*{Acknowledgments}

WOL acknowledges support from the National Science Foundation Graduate Research Fellowship under Grant No.\ DGE-2236414. Any opinions, findings, and conclusions or recommendations expressed in this material are those of the author(s) and do not necessarily reflect the views of the National Science Foundation. ZS acknowledges funding from NASA grant number 80NSSC24M0021. ZS and WOL acknowledge useful conversations with Mathias Garny, with members of the Slepian Group at the University of Florida, and during the online conference Parity Violation from Home.


\bibliographystyle{JHEP}
\bibliography{references}

\end{document}